\documentclass[epj-spec]{svjour}
\usepackage{graphics}
\begin{document}
\title{Entanglement and the Kondo effect in double quantum dots}
\author{Anton Ram{\v s}ak\inst{1,2}\fnmsep
\thanks{\email{anton.ramsak@fmf.uni-lj.si}} \and Jernej 
Mravlje\inst{2} }
\institute{Faculty of Mathematics and Physics, University of Ljubljana,
Slovenia \and Jo\v{z}ef Stefan Institute, Ljubljana, Slovenia }
\abstract{
We investigate entanglement between electrons in serially coupled
double quantum dots attached to non interacting leads. In addition to
local repulsion we consider the influence of capacitive inter-dot
interaction. We show how the competition between extended Kondo and
local singlet phases determines the ground state and thereby the
entanglement. }
% phases in spin and charge degrees of freedom

% We investigate pair entanglement of electrons and conductance in
% serially coupled double quantum dots attached to non interacting 
% leads.  The emphasis
% is on the numerical analysis of finite inter-dot tunneling in the
% presence of intra- and inter-dot repulsive coupling. The results
% reveal the competition between extended Kondo phases and local singlet
% phases in spin and charge degrees of freedom. 
 %end of abstract
%
\maketitle
\section{Introduction}
\label{intro}
In early days of quantum mechanics the entanglement between
particles was considered  a paradox. Today it 
has become appreciated that the ability to establish entanglement between
qubits in a controlled manner is a crucial ingredient of any quantum
information processing system. The interest in such systems is  spurred
on also by the fact that if a quantum computer were built, it would be
capable of tasks impracticable in classical computing \cite{nielsen01}
as are, \emph{e.g.}, factoring and searching algorithms  \cite{shor94}.

In general, it is desirable that the quantum computing hardware meets
several criteria as originally proposed by DiVincenzo \cite{criteria}
and include (i)  well defined qubits with the feasibility to scale up
in number; (ii) the possibility to initialize and manipulate qubit  
states; (iii) decoherence processes should be minimal that quantum
error correction techniques can be applied; and (iv) ability of
detecting final qubit states as the outcome of quantum computation. It
seems that these criteria for scalable qubits can be met in structures
consisting of coupled quantum dots \cite{divincenzo05,coish06} which
are therefore considered for implementation of quantum computing
processes in solid state. 

In particular, recent experiments on semiconductor double quantum dot (DQD)
devices have shown the evidence of spin
entangled states in GaAs based heterostuctures \cite{chen04}. 
It was shown that vertical-lateral double
quantum dots may be useful for implementing two-electron spin
entanglement \cite{hatano05} and it was demonstrated that coherent
manipulation and projective readout is possible in double quantum dot
systems \cite{petta05}. The ability to precisely control the number of electrons
by surface gates was also reported \cite{elzerman03}.

One of the central issues regarding two-qubit operations as
the basis for quantum computing algorithms is the creation and the
control of qubit pair entanglement in a computing device
\cite{nielsen01}. The interaction of
qubit pairs with their environment is in general a complicated
many-body process and its understanding is crucial for experimental
solid state realization of qubits in single and double quantum dots
\cite{coish06}.  

Specifically, the Kondo effect was found to play an important role in
single \cite{goldhaber-gordon98} and double quantum dot
\cite{heong01,wilhelm02,holleitner02} systems and here we report how
the Kondo interaction diminishes the entanglement between qubits defined
in DQDs even when other sources of decoherence (\emph{e.g.} phonons)
are absent.

\section{Coupled quantum dots with interaction}

\begin{figure}
\center{\resizebox{0.6\columnwidth}{!}{%
\includegraphics{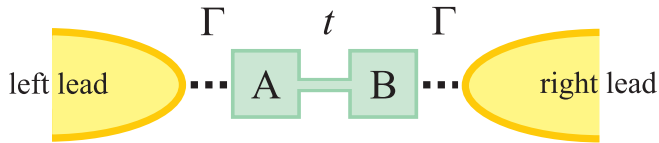} }}
\caption{Schematic picture of serial DQD coupled to leads.}
\label{Fig1}       % Give a unique label
\end{figure}

We consider a serially coupled DQD: a device with the
ability to produce entangled pairs that may be extracted using a
single-electron turnstile \cite{hu05}. We model DQD using the
two-impurity Anderson Hamiltonian  

\begin{equation}
H= 
\sum_{i=\mathrm{A},\mathrm{B}}(\epsilon n_{i}+Un_{i\uparrow}n_{i\downarrow})
+ V n_{\mathrm{A}}n_{\mathrm{B}}
-t\sum_{s}(c_{\mathrm{A}s}^{\dagger}c_{\mathrm{B}s} + h.c.), 
\label{2and}
\end{equation}
where $c^\dagger_{is}$ creates an electron with spin $s$ in the dot
$i=\mathrm{A}$ or $i=\mathrm{B}$ and $n_{is}=c^\dagger_{is}c_{is}$ is
the number operator. The on-site energies $\epsilon$ and the Hubbard
repulsion $U$ are taken equal for both dots. The dots are coupled to
the left and right noninteracting tight-binding leads with the chemical
potential set to the middle of the band of width $4t_0$.
Each of the dots is coupled to the adjacent lead by hopping $t'$ and
the corresponding hybridization width is
$\Gamma=(t')^2/t_0$. Schematically this setup is presented in
Fig.~\ref{Fig1}. The dots are coupled  capacitively by a
inter-dot repulsion term $V n_{\mathrm{A}}n_{\mathrm{B}}$.

In this paper we concentrate on the low temperature properties of
DQD system determined from the ground state. We expand the ground state
in the Sch\"{o}nhammer and Gunnarsson projection-operator
basis  \cite{schonhammer76,schonhammer84} $
|\Psi_{\lambda\lambda'}\rangle=P_{(\lambda i) }P_{(\lambda' j)}
\left|\tilde{0}\right\rangle $, which consists of projectors
$P_{\lambda i}$ where $i\in\{A,B\}$,--  \emph{e.g.},
$P_{(0i)}=\left(1-n_{i\uparrow}\right)\left(1-n_{i\downarrow}\right)$,
$P_{(1i)}=\sum_{\sigma}n_{i\sigma}\left(1-n_{i\bar{\sigma}}\right)$,
$P_{(2i)}=n_{i\uparrow}n_{i\downarrow}$ -- and additional operators
involving the operators in leads. We used up to $\sim 100$ additional
combinations of operators consisting of, for example,
$P_{(3ji)}=P_{(0i)}\widehat{v_j} P_{(1i)}$, where $\widehat{v_j}$ denotes
the tunneling to/from dot $i$ to the site $j$ in the lead. 
These operators are applied to the
state $\left|\tilde{0}\right\rangle$, which is the ground state of the
auxiliary noninteracting DQD Hamiltonian of the same form as $H$, but
with $U,V=0$, renormalized parameters $\epsilon, t,t'\to
\tilde{\epsilon},\tilde{t},\tilde{t}'$ and additional parameter
$\tilde{t}''$ which corresponds to hopping from left dot to right lead
and vice versa which although absent in the original Hamiltonian is
present in the effective Hamiltonian in some parameter regimes.

The starting point towards the understanding of the ground state of DQDs are
the filling properties of isolated  DQDs. The first electron is
added when $\epsilon=t$, and the second when 
$\epsilon=-t+J-[(U+V)-|U-V|]/2$, where
$J=[-|U-V|+\sqrt{(U-V)^{2}+16t^{2}}]/2$ is the difference 
between singlet and triplet energies. For $\epsilon+U/2+V=0$ DQD is doubly occupied,
$n=\langle n_{\mathrm{A}}+n_{\mathrm{B}}\rangle=2$, and the
ground state is $\frac{1}{\sqrt{2}}[\alpha
(|{\uparrow\downarrow}\rangle-|{\downarrow\uparrow}\rangle) +
\beta(|{20}\rangle-|{02}\rangle)]$, where $\alpha/\beta =
4t/(V-U+\sqrt{(U-V)^2+16t^2})$. Here we use notation
$|{\uparrow\downarrow}\rangle$
corresponding to spin-up and spin-down states on sites $\mathrm{A}$ and $\mathrm{B}$,
and $|{20}\rangle$ to double and zero occupancy of sites $\mathrm{A,B}$.
The range of $\epsilon$ where single occupation is favorable is
progressively diminished when $V\neq U$. For large $t$ or at (and
near) $V=U$ the molecular bonding and anti-bonding orbitals are formed
as is seen here from $\alpha \sim \beta$.

When DQDs are attached to the leads the low temperature physics is to
the large extent the same as that of the 
two-impurity Kondo problem studied by Jones, Varma and Wilkins two
decades ago \cite{jones88,jones89}. There two impurities form either two
Kondo singlets with delocalized electrons or bind into a local
spin-singlet state which is virtually decoupled from delocalized
electrons. The crossover between the regimes is determined by the
relative values of the exchange magnetic energy $J$ and twice the
Kondo condensation energy, of order the Kondo temperature given by the Haldane formula,
$T_K=\sqrt{U\Gamma/2}\exp(-\pi \epsilon(\epsilon+U)/2\Gamma)$. Such
results were obtained by the analysis of a two-impurity Anderson model
by means of slave-boson formalism
 \cite{aono98,georges99,aguado00,aono01,lopez02},  numerical
renormalization group \cite{izumida99,izumida00,rmzb06} or  
present formalism \cite{rmzb06,v2}.
Resembling
behavior was found also in particular regimes of triple quantum dot
systems \cite{zitko06}, and DQDs in  side coupled \cite{zitko06b} and parallel
\cite{zitko06c} configurations.

\section{Entanglement}
\label{sec:1}
\subsection{Spin entanglement}

Quantum entanglement as a physical resource was first defined for 
two distinguishable particles in a pure
state through von Neuman entropy and concurrence 
\cite{bennett96,hill97,vedral97,wootters98}. However, amongst the
realistic systems of major physical interest, electron-qubits have the
potential for a much richer variety of entanglement measure choices
due to both their charge and spin degrees of freedom.  In systems of
identical particles, for example, generalizations are needed to define
an appropriate entanglement measure which adequately deals with
multiple occupancy states \cite{schliemann01,ghirardi04,eckert02,gittings02}.

When entanglement is quantified in fermionic systems the measure must
also account for the effect of exchange  \cite{vedral03} as well as of
mutual electron repulsion.  In lattice fermion models 
entanglement is sensitive to the interplay between
charge hopping and the avoidance of double occupancy due to the Hubbard
repulsion, which results in an effective Heisenberg interaction
between adjacent spins  \cite{zanardi02}.  Entangled fermionic qubits
can be created with electron-hole pairs in a Fermi sea \cite{beenakker05} 
and in the scattering of two distinguishable
particles  \cite{bertoni03}. A spin-independent scheme for detecting
orbital entanglement of two-quasiparticle excitations of a mesoscopic
normal-superconductor system was also proposed recently \cite{samuelsson03}.

For two distinguishable particles A and B, described  with
single spin-$\frac{1}{2}$ (or pseudo spin) states $s=\uparrow$ or
$\downarrow$ and in a pure state 
$|\Psi_{\mathrm{AB}}\rangle=\sum_{ss'}\alpha_{ss'}|s\rangle_{\! \mathrm{A}}
|s'\rangle_{\! \mathrm{B}}$
concurrence as a measure of entanglement is given by  \cite{hill97}

\begin{equation}
C_0 = 2|\alpha_{\uparrow\!\uparrow}\alpha_{\downarrow\!\downarrow}-
\alpha_{\uparrow\!\downarrow}\alpha_{\downarrow\!\uparrow}|.\label{eq:wootters}\end{equation}
Two qubits are completely entangled, $C=1$,
if they are in one of the Bell states \cite{bennett96}, \emph{e.g.}, singlet
$|\Psi_{\mathrm{A}\mathrm{B}}\rangle \propto {|\uparrow
\downarrow\rangle -|\downarrow \uparrow \rangle}$.

A qubit pair represented by two electrons in DQDs and in the contact
with the leads acting as a fermionic bath can not be described by a
pure state and entanglement can not be related to the concurrence given
with the Wootters formula Eq.~(\ref{eq:wootters}). In the case of 
mixed states describing qubit pairs concurrence is related to the
reduced density matrix of the DQD subsystem
\cite{wootters98,osterloh02,syljuasen03}, where for systems that are
axially symmetric in spin space the concurrence may conveniently be
given in the closed form \cite{ramsak06},
\begin{eqnarray}
C_0 & = & \textrm{max}(0,C_{\uparrow\!\downarrow},C_{\parallel})
,\label{eq:cmax}\nonumber\\
C_{\uparrow\!\downarrow} & = & 2|\langle S_{\mathrm{A}}^{+}
S_{\mathrm{B}}^{-}\rangle|-2\sqrt{\langle P_{\mathrm{A}}^
{\uparrow}P_{\mathrm{B}}^{\uparrow}\rangle\langle P_{\mathrm{A}}^
{\downarrow}P_{\mathrm{B}}^{\downarrow}\rangle}, \\
C_{\parallel} & = & 2|\langle S_{\mathrm{A}}^{+}S_{\mathrm{B}}^{+}
\rangle|-2\sqrt{\langle P_{\mathrm{A}}^{\uparrow}P_{\mathrm{B}}^
{\downarrow}\rangle\langle P_{\mathrm{A}}^{\downarrow}
P_{\mathrm{B}}^{\uparrow}\rangle},\nonumber \end{eqnarray}
where $S_i^{+} = (S_i^{-})^{\dagger} = c_{i\uparrow}^{\dagger}
c_{i\downarrow}$ is the electron spin raising operator for dot
$i=\mathrm{A}$ or $\mathrm{B}$ and $P_i^{s}=n_{is}(1-n_{i,-s})$ is the
projection operator onto the subspace where dot $i$ is occupied by one
electron with spin $s$. 

In the derivation of concurrence formula Eq.~(\ref{eq:cmax}) the
reduced density matrix was obtained by projecting onto four local spin
states of $|\!\uparrow\,\rangle_\mathrm{A}$,
$|\!\downarrow\,\rangle_\mathrm{A}$,
$|\!\uparrow\,\rangle_\mathrm{B}$, and
$|\!\downarrow\,\rangle_\mathrm{B}$, corresponding to {\it singly}
occupied DQD sites A and B, respectively.  If $t/U$ is not small the
electrons tunnel between the dots and charge fluctuations introduce
additional states with zero or double occupancy of individual dots
\cite{schliemann01,zanardi02}.  As pointed out by Zanardi
\cite{zanardi02} in the case of simple Hubbard dimer the entanglement
is not related only to spin but also to charge degrees of freedom
which emerge when repulsion between electrons is weak or moderate.

For systems with strong electron-electron repulsion, charge
fluctuations are suppressed and the states with single occupancy --
the spin-qubits -- dominate: the concept of spin-entanglement
quantified with concurrence can still be applied.  We use
spin-projected density matrix and consider only entanglement
corresponding to {\it spin} degrees of freedom. Due to doubly (or
zero) occupied states arising from charge
fluctuation on the dots (caused by tunneling between the dots A and B or
due to the exchange with the electrons in the leads), the reduced
density matrix has to be renormalized. The probability that at the
measurement of entanglement there is precisely one electron on each of
the dots is less than unity, $P_{11}<1$, and the spin-concurrence is
then given with

\begin{equation}
C=C_0/P_{11},\label{cnorm}
\end{equation}
where $P_{11}=P_{\uparrow\downarrow}+P_{\parallel}$, and
$P_{\uparrow\downarrow} =\langle 
P^\uparrow_\mathrm{A} P^\downarrow_\mathrm{B} +
P^\downarrow_\mathrm{A} P^\uparrow_\mathrm{B}\rangle $,
$P_{\parallel} = 
\langle P^\uparrow_\mathrm{A} P^\uparrow_\mathrm{B} + P^\downarrow_\mathrm{A}
P^\downarrow_\mathrm{B}\rangle$ are probabilities for antiparallel and parallel
spin alignment, respectively. Such procedure corresponds to the
measurement apparatus which would only discern spins and ignore
all cases whenever no electron, or a electron pair would appear at one
of the detectors at sites A or B.

\subsection{Charge (isospin) entanglement}

In the ground state of two isolated impurities coupled by
a capacitive (but not tunneling) term $V=U$,  4 'spin
states' $|{\sigma_1 \sigma_2}\rangle$ and 
2 'charge states' $|{20}\rangle$, $|{02}\rangle$ are
degenerate. By introducing the pseudospin operator
 \cite{leo04} $\tilde{T}^i= 1/2\sum_{ll'=1,2}\sum_\sigma
c^{\dagger}_{l\sigma}\tau^i_{ll'}c_{l'\sigma} $, where $\tau^i$ are
the Pauli matrices, and the combined spin-pseudospin operators
$W^{ij}=S^i \tilde{T}^j$, the Hamiltonian is SU(4)
symmetric. As long as the SU(4) symmetry breaking terms are small
enough (\emph{e.g.}, tunneling $t\to 0$) the ground state of such DQDs attached to the leads
remains close to an SU(4) state with
'spin' screened by the electrons in the leads  \cite{v2}.  

If $V >\!\!> U$ charge states dominate and in this case {\it charge
concurrence} can be defined in a direct analogy with the previous spin
case. In Eq.~(\ref{eq:cmax}) one just has to replace the spin operators with their
corresponding isospin counterparts, \emph{e.g.},
$S_{\lambda}^{-}=c_{\lambda\downarrow}^{\dagger}c_{\lambda\uparrow}=(S_{\lambda}^{+})^{\dagger}$
$\to$
$T_{\lambda}^{-}=c_{\lambda\uparrow}c_{\lambda\downarrow}=(T_{\lambda}^{+})^{\dagger}$
for sites $\lambda=$A,B and
$S_{\lambda}^{z}=(n_{\lambda\uparrow}-n_{\lambda\downarrow})/2$ $\to$
$T_{\lambda}^{z}=(n_{\lambda}-1)/2$. If the probability for spin
states is significant, appropriate renormalization to charge states is
analogous to Eq.~(\ref{cnorm}), but with corresponding isospin
operators. The density matrix is here renormalized with the probability $P_{20}$ that
precisely $n_{\mathrm{A},\mathrm{B}}=2$ and $n_{\mathrm{B},\mathrm{A}}=0$ 
electrons occupy individual dots (corresponding to apparatus which
only measures occurrence or absence of pairs at each of its detectors).

\section{Numerical results}
\subsection{Concurrence}

We here present the results for zero temperature 
(ground state) concurrence of a qubit pair in DQDs
in the absence of magnetic field. Temperature dependence of 
concurrence for the case of $V=0$ is
given in Ref.~\cite{rmzb06}.
Expectation values $\langle ...\rangle$ in
the concurrence formula Eq.~(\ref{eq:cmax}) are now calculated using the ground
state therefore $\langle
S_{\mathrm{A}}^{+}S_{\mathrm{B}}^{+}\rangle=0$ and $C_{\parallel}<0$.  
We consider the particle-hole symmetry point
with $n=2$ and $\epsilon+U/2+V=0$.

\begin{figure}
\center{\resizebox{0.55\columnwidth}{!}{%
\includegraphics{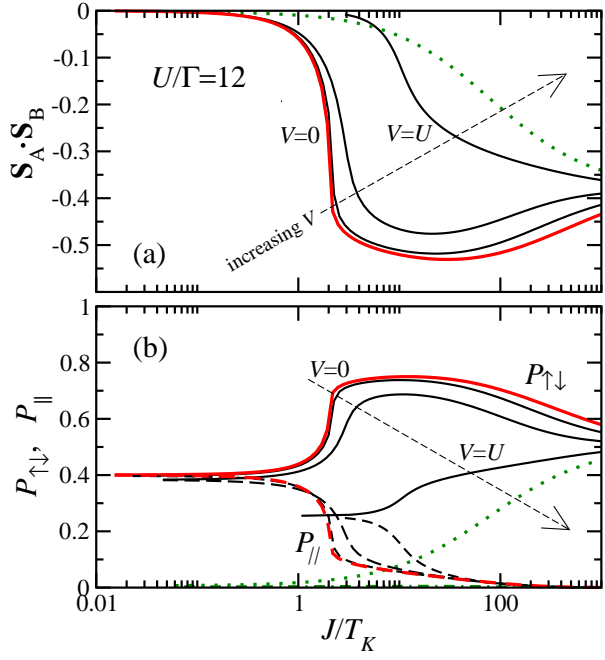} }}
\caption{(a) Spin-spin correlation  for $V/U=0,1/3,2/3,1$ (full
  lines)  and $V/U=5/4$ (dotted) for
$U/\Gamma=12, \Gamma/t_0=0.1$. (b) Probabilities for  parallel 
(lower curves - dashed) and anti-parallel
  (upper curves - full) spins of electrons in the DQD for $V/U$ ratios as in (a). 
Note that the probability 
  for parallel spins for $V/U=5/4$ is almost zero (dashed-dotted), while
$P_{\uparrow\downarrow}<1/2$ for $J/T_K<1000$ (dotted); the probabilities do not sum
  to 1. The deficiency (which goes to zero as $U\to \infty$) is due to
  states with double particle (or hole) 
  occupancy on at least one dot. }
\label{Fig2}       % Give a unique label
\end{figure}

\begin{figure}
\center{\resizebox{0.55\columnwidth}{!}{%
\includegraphics{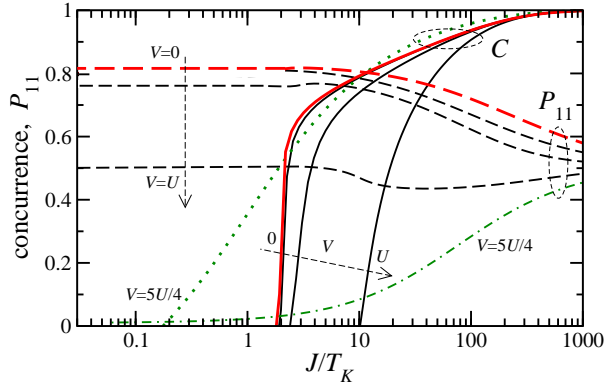} }}
\caption{Concurrence (full curves) and single particular dot occupation probability
$P_{11}$ (dashed) for
  $V/U=0,1/3,2/3,1$. For $V/U=5/4$ the concurrence and $P_{11}$ are
  plotted dotted and dashed-dotted, respectively. Parameters are as in Fig.~\ref{Fig2}.}
\label{Fig3}       % Give a unique label
\end{figure}

Qualitatively, the concurrence
is significant whenever enhanced spin-spin correlations indicate inter-dot
singlet formation. As shown in Fig.~\ref{Fig2}(a) for $U/\Gamma=12$ and
$\Gamma/t_0=0.1$, the correlation function
$\langle {\bf S}_\mathrm{A} \cdot {\bf S}_\mathrm{B} \rangle$ 
tends to $-3/4$ for $J$ large enough to suppress
the formation of Kondo singlets, but still  $J/U\ll1$, that
local charge fluctuations are sufficiently suppressed. 
In particular, the local dot-dot singlet is formed whenever
singlet-triplet splitting superexchange energy $J>J_{c}\sim 2T_{K}$.
With increasing
$V\to U$, and above $U$, the probability for singly occupied spin states,
$P_{11}=P_{\uparrow\downarrow}+P_{\parallel}$
is significantly reduced, Fig.~\ref{Fig2}(b) and Fig.~\ref{Fig3},
which also leads to reduced spin-spin corelation, Fig.~\ref{Fig2}(a).
In this limit the concept involving isospin entanglement can be applied
(not shown here).

Concurrence, corresponding to the correlation function from Figs.~\ref{Fig2}(a,b) is
presented in Fig.~\ref{Fig3} for various values of $V$. 
As discussed above, $C$ is zero for $J$ below $\sim 2T_K$ due to the Kondo effect,
which leads to entanglement between localized and conducting electrons \cite{adam06}
instead of the $\mathrm{A}$-$\mathrm{B}$ qubit pair entanglement. In finite magnetic field
irrespectively of temperature the 
concurrence abruptly tends to zero for $B>J$ (not shown here) \cite{magnetic}.

\subsection{Conductance}

One of the most directly  measurable properties of DQDs is the linear
conductance. We calculate the zero temperature conductance  
using the sine formula (SF)
 \cite{rr03,rr03b,rbr04}, $G=G_{0}\sin^{2}[(E_{+}-E_{-})/4t_{0}L]$, where
$G_{0}=2e^{2}/h$ and $E_{\pm}$ are the ground state energies of a
large auxiliary ring consisting of $L$ non-interacting sites and an
embedded DQD, with periodic and anti-periodic boundary conditions,
respectively. Alternatively conductance can be obtained also from the
Green's function (GF) corresponding to the effective noninteracting
Hamiltonian $\tilde{H}$ with effective parameters \cite{rr03,cm}.  The
advantage of the former method is better convergence in the strong
coupling regime, however, the accuracy of the SF method depends only
on the accuracy of the ground-state energy and is therefore in some
cases more robust. By comparing results of both methods we checked for
the consistency and the convergence.

\begin{figure}
\center{\resizebox{0.75\columnwidth}{!}{%
\includegraphics{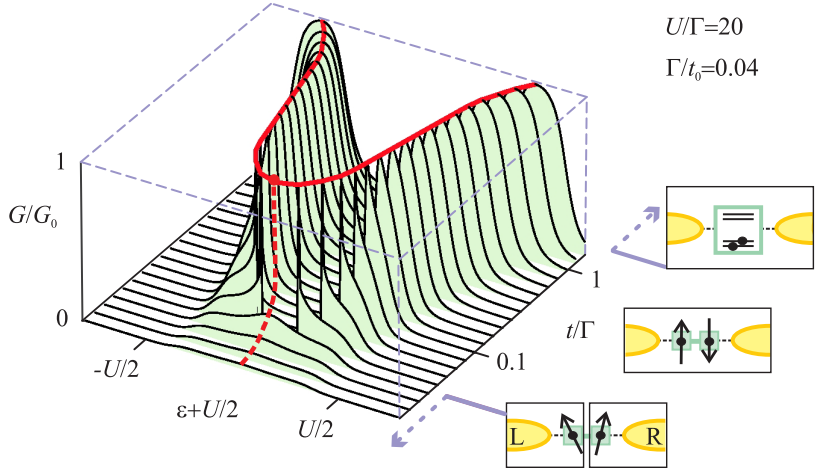} }}
\caption{Conductance of DQD as a function of gate voltage and
  inter-dot tunneling rate for  $U/\Gamma=20, \Gamma/t_0=0.04$. 
Pictograms indicate dominant ground state regimes: molecular orbital Kondo effect, 
local spin-singlet formation and
two separate Kondo effects.}
\label{Fig4}       % Give a unique label
\end{figure}

The conductance as a function of interdot hoping $t$ and
$\epsilon+U/2$ in the absence of interdot repulsion, $V=0$, is
presented in Fig.~\ref{Fig4}. The Hubbard repulsion is set to
$U/\Gamma=20$ and hibridization to $\Gamma/t_0=0.04$. As in the previous Section
three different regimes of $t$ correspondingly reflect 
in the results for conductance. For large
$t/\Gamma>1$ (but with $\Gamma/U <\!\!< 1$) the DQD is in
molecular-orbital Kondo regime when occupancy is odd, i.e,
$n\sim1,3$. Typical Kondo conductance plateau of width $\sim U/2$ is
developed around bonding (and anti-bonding) molecular orbital level
$\pm t$. These two unitary conductance regions with reducing $t$
become progressively sharper when we enter $t/U <\!\!< 1$
regime. There the description of DQD in terms of bonding/anti-bonding
orbitals should be replaced with local picture. Due to strong
electron-electron repulsion local charge fluctuations are
suppressed  at the point of particle-hole symmetry with $n=2$ and
$\epsilon+U/2=0$. Thick full line corresponds to points of $G=G_0$ and there
exists some critical $t_c$ where two conductance peaks merge (bullet) and for $t<t_c$
conductance is less then $G_0$ (dashed line). 
The corresponding critical superexchange interaction is of the order of Kondo temperature
as before, $J_c\sim 2 T_K$. 
As discussed above, in this regime each of the dots undergoes the Kondo effect
where local moment is screened by conducting electrons in the adjacent lead and left --
right sides of the system become decoupled which leads to vanishing  A$\to$B  conductance
as $G\propto (t/\Gamma)^2$  \cite{aono98}.

\section{Summary}

\begin{figure}
\center{\resizebox{0.9\columnwidth}{!}{%
\includegraphics{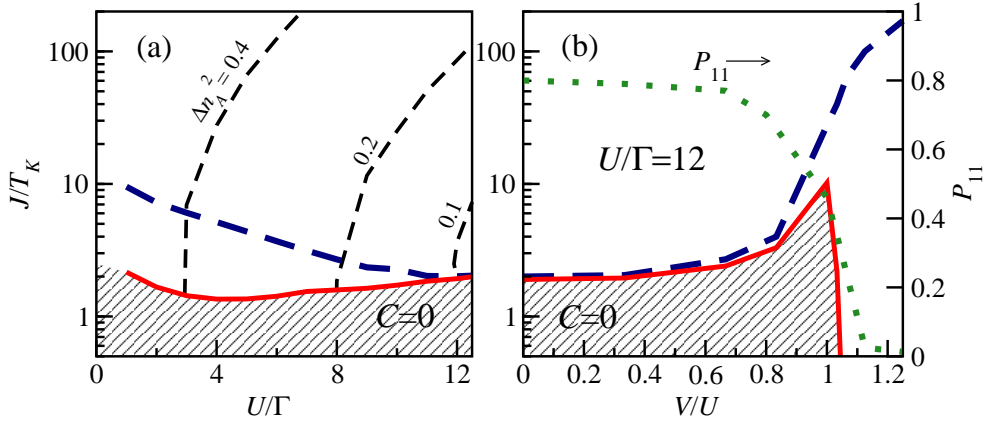} }}
\caption{ (a) Charge fluctuations (short-dashed), $\textrm{\bf S}_A\cdot
  \textrm{\bf S}_B=-1/4$ (long-dashed) and $C=0$ (full) in the $(U/\Gamma,J/T_K)$ 
plane. (b) $\textrm{\bf S}_A\cdot
  \textrm{\bf S}_B=-1/4$ (long-dashed) and $C=0$ (full) in the $(V/U,J/T_K)$ 
plane. $P_{11}$ is shown with dotted line (right scale).}
\label{Fig5}       % Give a unique label
\end{figure}

The main results concerning entanglement of qubit pairs
in serially coupled double quantum dots are extracted in Fig.~\ref{Fig5}.
The charge
fluctuations $\Delta n_\mathrm{A}^2=\langle n_\mathrm{A}^2\rangle-\langle n_\mathrm{A} \rangle^2$, 
contour plot in  Figs.~\ref{Fig5}(a), are suppressed for
sufficiently large repulsion, \emph{e.g.}, $U/\Gamma > 10$. In this limit and
in vanishing magnetic field, the DQD can be described in terms of the
Werner states \cite{werner89} and becomes similar to recently studied
problem of entanglement of two Kondo spin impurities embedded in a
conduction band \cite{cho06}. In this case,
$C_{\uparrow \downarrow}\sim 2(-\langle\textrm{\bf
S}_{\mathrm{A}}\cdot\textrm{\bf
S}_{\mathrm{B}}\rangle-\frac{1}{4})\sim P_{\uparrow
\downarrow}-2P_{||}$ for $C_{\uparrow \downarrow}\geq 0$. For large
$U/\Gamma$, where the charge fluctuations vanish, the
$\langle\textrm{\bf S}_{\mathrm{A}}\cdot\textrm{\bf
S}_{\mathrm{B}}\rangle=-\frac{1}{4}$ line (dashed
line) progressively merges with the $C=0$ boundary line (full).

In Fig.~\ref{Fig5}(b) phase diagram with fixed $U/\Gamma=12$ and corresponding to
Fig.~\ref{Fig5}(a) presents $V/U$ dependence of $C=0$ boundary line (full) in comparison with 
the $\langle\textrm{\bf S}_{\mathrm{A}}\cdot\textrm{\bf
S}_{\mathrm{B}}\rangle=-\frac{1}{4}$ line (dashed). With $V$ exceeding $U$ the probability
for well defined spin-qubit pairs in DQD, $P_{11}$, 
rapidly decreases which means that states with doubly  occupied or empty 
individual dots dominate.
In this regime $C=0$ line is pushed to much lower $J/T_K$. For $V>U$ the probability 
$P_{11}$ becomes progressively negligible giving more meaning to
considering charge (isospin) entanglement instead. It should be noted,
however, that in realistic DQD systems 
intersite repulsion $V$ is in general weaker compared to $U$ and
that this regime would not be easily reached experimentally. One
possibility, where $V$-interaction could dominate, are systems with
strong local electron-phonon interactions which may significantly
renormalize local $U$  \cite{jm05} without affecting capacitive
interaction $V$.

To conclude, we have found generic behavior of spin-entanglement of an electron
pair in serially coupled double quantum dots. On the one hand, we have shown
quantitatively that making the spin-spin exchange coupling $J$ large
by increasing tunneling $t$, leads to enhanced charge fluctuations,
whilst on the other, at small magnetic interactions $J<J_{c}$ 
entanglement is suppressed as the DQD system undergoes  the Kondo
effect.  Various regimes are explained  and supported with
typical numerical examples.

We thank T. Rejec for his GS code and suggestions. We acknowledge 
J. Bon{\v c}a and R. {\v Z}itko  for useful discussions.
We also acknowledge support from the Slovenian Research
Agency under contract Pl-0044.

\end{document}